\documentclass[9pt,twoside]{pnas-new}

\templatetype{pnasmathematics} 

\usepackage{subfigure}
\usepackage{wrapfig}
\newcommand{\bm}[1]{\mbox{\boldmath$#1$}}

\title{A Variational Theory of Lift}

\author[a]{Cody Gonzalez}
\author[a,2]{Haithem E Taha} 

\affil[a]{University  of  California,  Irvine}

\significancestatement{After more than a century since the first successful flight, we managed to discover the general mathematical principle that governs lift generation over airfoils; we developed a new theory of lift that dispenses with the ubiquitously used empirical condition---aka the Kutta condition. Interestingly, the developed theory is based on a principle due to Hertz that seems to have gone into oblivion in the history of mechanics: Hertz’ principle of least curvature. The developed (inviscid) principle reduces to the Kutta condition in the special case of a sharp-edged airfoil, which challenges the widely accepted wisdom about the viscous nature of the Kutta condition. Moreover, the developed theory, unlike the classical theory, allows computing lift over smooth shapes (without sharp edges).}

\authorcontributions{H.T. introduced the use of variational principles in the airfoil problem; C.G. conjectured the minimality of curvature; H.T. drew the connection with Gauss' principle; C.G. reduced it to Hertz' principle and conjectured the constraint nature of the pressure force; H.T. proved it; C.G. performed numerical simulations; H.T. wrote the draft of the paper; and both authors finalized the manuscript.}
\authordeclaration{The authors declare no competing interests.}
\correspondingauthor{\textsuperscript{2}To whom correspondence should be addressed. E-mail: hetaha@uci.edu}


\begin{abstract}
In this paper, we revive a special, less-common, variational principle in analytical mechanics (Hertz’ principle of least curvature) to develop a novel variational analogue of Euler’s equations for the dynamics of an ideal fluid. The new variational formulation is fundamentally different from those formulations based on Hamilton's principle of least action. Using this new variational formulation,  we generalize the century-old problem of the flow over a two-dimensional body, to find that lift is a direct consequence of curvature. The developed variational principle reduces to the classical Kutta-Zhukovsky condition in the special case of a sharp-edged airfoil, which challenges the accepted wisdom about the Kutta condition being a manifestation of viscous effects.  Rather, we found that it represents conservation of momentum. Moreover, the developed variational principle provides, for the first time, a theoretical model for lift over smooth shapes without sharp edges where the Kutta condition is not applicable. We discuss how this fundamental divergence from current theory can explain discrepancies in computational studies and experiments with superfluids.
\end{abstract}

\dates{This manuscript was compiled on \today}
\doi{\url{www.pnas.org/cgi/doi/10.1073/pnas.XXXXXXXXXX}}

\begin{document}

\maketitle
\thispagestyle{firststyle}
\ifthenelse{\boolean{shortarticle}}{\ifthenelse{\boolean{singlecolumn}}{\abscontentformatted}{\abscontent}}{}

\section{A Meager State of Theoretical Modeling of Aerodynamics}
The problem of the flow over a lifting airfoil is a century-old, classical textbook problem in aerodynamics and fluid mechanics \cite{Schlichting,Karamcheti,Milne_Thomson_5th}. The problem is analytically solvable thanks to three elements. First, the potential flow around a circular cylinder is readily known since the 1877 seminal paper of Lord Rayleigh \cite{Rayleigh_Circular_Cylinder}. Second, the Riemann mapping theorem ensures that any simply connected domain can be (biholomorphically) mapped to the open disc. So, the flow around any two-dimensional shape can be easily constructed from the cylinder flow via conformal mapping between the cylinder and the shape of interest. However, this solution is not unique. One can always add a circulation of arbitrary strength at the center of the cylinder, which does not affect the no-penetration boundary condition at all. Interestingly, this circulation is of paramount importance for lift production; in fact, it solely dictates the amount of lift generated. Therefore, the potential-flow theory alone cannot predict the generated lift force; a closure condition must be provided to fix the dynamically-correct amount of circulation. The third element is the Kutta-Zhukovsky condition, which has traditionally provided such a closure.

The Kutta condition is quite intuitive: if the body has a sharp trailing edge leading to a singularity, then the circulation must be set to remove such a singularity; it is a singularity removal condition \cite{Kutta_Crighton}. It is quite accurate for a \textit{steady} flow (at a high Reynolds number and a small angle of attack);  indeed, it is a paradigm for engineering ingenuity where an enabling mathematical condition is inferred from physical observations.

Interestingly, if one wishes to consider a smooth trailing edge, however small the trailing radius (i.e., however close to a sharp trailing edge), the classical aerodynamic theory collapses; there are no theoretical models that can predict lift on a two-dimensional smooth body without sharp edges! (only few ad-hoc methods with no theoretical basis). In fact, some authors even consider the sharp edge as a lifting mechanism; i.e., an airfoil must have a sharp trailing edge to generate lift, see Ref. \cite{New_Theory_of_Flight}. 

For unsteady flows, the status of current aerodynamic theory is even worse. Martin Kutta never claimed that his condition works for unsteady flows; in fact, for an unsteady case, it is already known that, in the early transient period after an impulsive start, the flow goes around the trailing edge from the lower surface to the upper surface (see \cite{Prandtl_Lectures_Fundamentals_Applied}, pp. 158-168; \cite{Goldstein_Book1}, pp. 26-36; or \cite{Schlichting}, pp. 33-35). That is, since more than a century, it is known that the Kutta condition is not generally applicable to unsteady flows (even for airfoils with sharp trailing edges at small angles of attack). This is why numerous research reports criticized the application of the Kutta condition to unsteady flows \cite{Kutta_Failure_Zero_Alpha,Flutter_Failure1,Flutter_Failure2,Kutta_Failure_Zero_Lift,Oseen_Chu,Flutter_Failure3,Oseen_JFM,Kutta_Failure2,Kutta_Failure3,Kutta_Applicability_JFM,Kutta_Applicability1,Kutta_Applicability2,Kutta_Failure4,Kutta_Applicability3,Kutta_Crighton}. Nevertheless, almost all unsteady models, starting with the pioneering efforts in 1920's and 1930's of Wagner \cite{Wagner}, Theodorsen \cite{Theodorsen}, and Von Karman and Sears \cite{VonKarman_Sears} until the recent work of Badoo et al. \cite{Jaworsky_JFM_Porpous}, adopted the Kutta condition. There is no legitimate alternative! So, what we have is a meager state of knowledge and a very confined capability of aerodynamic theoretical modeling: we can only analyze steady flow at a small angle on a body with a sharp trailing edge! Basically, the aerodynamic theory is encumbered with the Kutta condition; it is applicable whenever the latter is befitting. It is clear that we lack a general theory for lift computation---equivalently a general closure principle, alternative to Kutta's, for potential flow. 
The above discussion is intimately related to the lifting mechanism. The accepted wisdom by the fluid mechanics community asserts that the Kutta condition is a manifestation of viscous effects---it implicitly accounts for viscous effects in a potential-flow formulation. In fact, this view has been challenged by several authors \cite{Lift_Quest,New_Theory_of_Flight}, but without providing a mathematical proof or theoretical basis. It may be the right time to recall Chang's 2003 New York Times article: \textit{What Does Keep Them Up There?} \cite{Lift_NYTimes}: ”\textit{To those who fear ﬂying, it is probably disconcerting that physicists and aeronautical engineers still passionately debate the fundamental issue underlying this endeavour: what keeps planes in the air?}".

In this paper, we revive a special variational principle from the history of analytical mechanics: Hertz' principle of least curvature. Exploiting such a variational principle, we develop a novel variational formulation of Euler's equations for the dynamics of ideal fluids where the \textit{Appellian} (integral of squared acceleration) is minimized. The new variational formulation is fundamentally different from the several developments based on Hamilton's principle of least action \cite{Variational_Principles_Fluids1,Bateman_Variational_Principle,Variational_Principles_Fluids2,Variational_Principles_Fluids25,Variational_Principles_Fluids_Hamilton_PoF,Variational_Principles_Fluids4,Seliger_Variational_Principles,Variational_Principles_Fluids5,Variational_Principles_Fluids_Hamiltonian_Ann_Rev,Variational_Principles_Fluids3}. It even transcends Euler's equation in some singular cases where the latter does not provide a unique solution, such as the airfoil problem.

Applying the developed variational formulation to the century-old problem of the flow over an airfoil, we develop a new variational theory of lift that dispenses with the Kutta condition. The Hertz principle of least curvature is used, as a first principle, to develop, for the first time, a closure condition alternative to Kutta's for the potential flow over an airfoil. In contrast to the Kutta condition, the developed closure condition is derived from first principles. The new variational condition reduces to the Kutta condition in the special case of a sharp-edged airfoil, which challenges the accepted wisdom about the Kutta condition being a manifestation of viscous effects. Rather, it is found that the Kutta condition is a momentum conservation mechanism for the inviscid flow over a sharp-edged airfoil. Moreover, the developed theory, unlike the classical theory, allows treatment of smooth, singularity-free cylinders. 


\section{Lack of Dynamical Features in the Current Theory of Aerodynamics: Variational Formulation is the Solution}
To solve for the flow field of an incompressible fluid, both the continuity (kinematics) and momentum (dynamics) equations must be solved simultaneously. However, in potential flow, the governing equation is the Laplacian in the velocity potential ($\bm\nabla^2\phi=0$), which is obtained by combining the continuity equation (a divergence-free constraint: $\bm\nabla\cdot\bm{u}=0$) with an irrotational-flow assumption (a curl-free constraint: $\bm\nabla\times\bm{u}=0$). These are kinematic constraints on the velocity field $\bm{u}$. Then, the momentum equation (Bernoulli's in this case) is considered next to solve for the pressure field. That is, in potential flow, the flow kinematics is decoupled from the flow dynamics; the velocity field is determined from purely kinematic analysis without any consideration for dynamical aspects. Therefore, it is fair to expect that such a pure kinematic analysis is not sufficient to uniquely determine the flow field \footnote{This is not the case for three-dimensional flows because the domain enclosing a 3D surface is simply connected while that around a 2D object is doubly connected: a closed circuit around the 3D surface is a reducible circuit while that around a 2D object is irreducible; there is no means to determine the circulation of an irreducible circuit.}; the fix must come from a dynamical consideration. 


Based on the above discussion, a proper closure condition (i.e., a condition that provides circulation dynamics) in potential flow must come from dynamical considerations. The challenge is: Can we project Euler's dynamical equations on a one-dimensional manifold to extract the dynamics of circulation alone? 

Dynamical equations of motion can be determined either from a Newtonian mechanics perspective or an analytical mechanics one. 
The former stipulates isolating fluid particles and writing the equations of motion for each individual particle even if the free variables in the system are significantly fewer than the total number of spatial coordinates of all individual particles due to kinematic or geometric constraints. On the other hand, the analytical (Lagrangian or variational) mechanics approach allows accepting kinematical constraints, ignoring the unknown forces that maintain them, and hence focusing on the relevant equations of motion; it provides directly the relevant equations of motion for the free variables by minimizing a certain objective function.

Projecting this discussion on the potential-flow case, one finds that the kinematical constraints of potential flow allow construction of the entire flow field $\bm{u}$ from the circulation free variable $\Gamma$ only: $\bm{u}=\bm{u}(\bm{x};\Gamma)$. That is, while there are infinite degrees of freedom for the infinite fluid particles, there is only one free variable (the circulation) which, via the potential-flow kinematical constraints, can be used to recover the motion of these infinite degrees of freedom. Hence, the analytical/variational mechanics appears to be specially well-suited for this problem; it will provide a single equation for the unknown circulation without paying attention to the irrelevant degrees of freedom of the fluid particles or the unknown forces that maintain kinematical constraints. Simply, the first variation of the ``objective function" with respect to circulation must vanish---and this necessary condition should provide a single dynamical equation in the unknown circulation.

Based on the above discussion, two important conclusions are drawn: (i) A true closure/auxiliary condition for potential flow must come from dynamical considerations; and (ii) Variational principles would be particularly useful to derive such dynamics. Insofar this story looks appealing, the question has always been: What is a suitable variational principle for the airfoil problem? \textbf{What is the special quantity that is being minimized in every flow over a two-dimensional body?}

\section{Theoretical Mechanics Approach: Gauss' Principle of Least Constraint and Hertz' Principle of Least Curvature}
There have been several variational formulations of Euler's equations; almost all of them are based on Hamilton's principle of least action \cite{Variational_Principles_Fluids1,Bateman_Variational_Principle,Variational_Principles_Fluids2,Variational_Principles_Fluids25,Variational_Principles_Fluids_Hamilton_PoF,Variational_Principles_Fluids4,Seliger_Variational_Principles,Variational_Principles_Fluids5,Variational_Principles_Fluids_Hamiltonian_Ann_Rev,Variational_Principles_Fluids3}. However, Hamilton's principle suffer from some drawbacks that precludes its applicability to the airfoil problem. In particular, Hamilton's principle is a \textit{time-integral} variational principle. That is, it provides the dynamics over a period of time; hence, it may not be applicable to a steady snapshot of a flow field. Also, Hamilton's principle is exactly equivalent to Newton's equations, so it does not provide more information in the singular cases where Newtonian (and Lagrangian) mechanics fail to determine a unique solution.

In fact, the search for a suitable variational formulation for the airfoil problem is challenging. For example, minimizing the kinetic energy over the field yields trivial (zero circulation) at any angle of attack. We found that the deserted principle of least constraint by Gauss provides a felicitous formulation for the current problem \footnote{Papastavridis wrote: ``In most of the $20^{\rm{th}}$ century English literature, GP [Gauss Principle] has been barely tolerated as a clever but essentially useless academic curiosity, when it was mentioned at all"\cite{Papastavridis}}.

Consider the dynamics of $N$ particles, each of mass $m_i$, which are governed by Newton's equations
\begin{equation}\label{eq:Newton}
  m_i \bm{a}_i = \bm{F}_i + \bm{R}_i \; \forall i\in\{1,..,N\},
\end{equation}
where $\bm{a}_i$ is the inertial acceleration of the $i^{\rm{th}}$ particle, and the right hand side represents the total force acting on the particle, which is typically decomposed in analytical mechanics into:  (i) impressed forces $\bm{F}_i$, which are the directly applied (driving) forces (e.g., gravity, elastic, viscous); and (ii) constraint forces $\bm{R_i}$ whose raison d'etre is to maintain/satisfy kinematical/geometrical constraints; they are passive or workless forces \cite{Lanczos_Variational_Mechanics_Book} \footnote{For time-varying constraints, only the virtual work of the constraint forces vanishes, but not necessarily their actual work \cite[pp. 382-383]{Papastavridis}}. That is, they do not contribute to the motion abiding by the constraint; their sole role is to preserve the constraint (i.e., prevent any deviation from it).

Inspired by his method of least squares, Gauss asserted that the deviation of the actual motion $\bm{a}$ from the impressed one $\frac{\bm{F}}{m}$ is minimum \cite{Gauss_Least_Constraint}. That is, the quantity
\begin{equation}\label{eq:Gauss}
Z = \sum_{i=1}^N \frac{1}{2} m_i\left( \frac{\bm{F}_i}{m_i} - \bm{a}_i \right)^2
\end{equation}
is minimum \cite[pp. 911-912]{Papastavridis}. Several points are worthy of clarification here. First, Gauss principle is equivalent to (derivable from) Lagrange's equations of motion \cite[pp. 913-925]{Papastavridis}, so we emphasize that it bears the same truth and status of first principles (Newton's equations). Second, in Gauss' principle, $J$ is actually minimum, not just stationary. Third, unlike the time-integral principle of least action, Gauss' principle is applied instantaneously (at each point in time). So, it can be applied to a particular snapshot. Fourth, in some of the singular cases (when the dimension of the tangent space changes),  Newtonian and Lagrangian mechanics (as well as the principle of least action) may fail to determine a unique solution. In contrast, Gauss' principle is capable of providing a unique solution, see \cite{Papastavridis}, pp. 923; and \cite{Golomb}, pp. 71-72.

In the case of no impressed forces
\[ m_i \bm{a}_i = \bm{R}_i \; \forall i\in\{1,..,N\}, \]
Gauss' principle reduces to Hertz' principle of least curvature, which states that the \textit{Appellian}
\begin{equation}\label{eq:Hertz}
  S = \sum_{i=1}^N \frac{1}{2}m_i\bm{a}_i^2
\end{equation}
is minimum. In this case, because kinetic energy is conserved, it can be shown that the system curvature is minimum \cite[pp. 930-932]{Papastavridis}. That is, a free (unforced) particle moves along a straight line. If it is a constrained motion, then it will deviate from a straight line to satisfy the constraint, but the deviation from the straight line path (i.e., curvature) would be minimum.

\section{Novel Variational Formulation of the Dynamics of Ideal Fluids}
Recall the Euler equations for incompressible flows
\begin{align}
\label{eq:Euler_eq}
\rho\bm{a} =& -\bm\nabla p, \mathrm{ \hspace{15pt} in  \hspace{15pt} \Omega}
\intertext{subject to continuity}
\label{eq:Euler_cont}
\bm\nabla \cdot \bm{u} =& 0, \mathrm{ \hspace{15pt} in  \hspace{15pt} \Omega}\\
\intertext{and the no-penetration boundary condition}
\label{eq:Euler_BC}
\bm{u}\cdot \bm{n} =& 0, \mathrm{ \hspace{15pt} on  \hspace{15pt} \partial{\Omega}},
\end{align}
where $\Omega$ is the spatial domain, $\partial\Omega$ is its boundary, $\bm{n}$ is normal to the boundary, and $\bm{a}=\frac{\partial \bm{u}}{\partial t}+\bm{u}\cdot\bm\nabla \bm{u}$ is the total acceleration of a fluid particle.

Equation [\ref{eq:Euler_eq}] presents Newton's equations of motion for the fluid parcels. For inviscid flows, neglecting gravity, the only acting force on a fluid parcel is the pressure force $\bm\nabla p$. In order to apply Gauss' principle, we must determine whether this force is an impressed force or a constraint force. Interestingly, for incompressible flows, it is the latter. The sole role of the pressure force in incompressible flows is to maintain the continuity constraint: the divergence-free kinematic constraint on the velocity field ($\bm\nabla \cdot \bm{u} = 0$). It is straightforward to show that if $\bm{u}$ satisfies Eqs. [\ref{eq:Euler_cont},\ref{eq:Euler_BC}], then \cite[][pp. 261]{Geometric_Control_Fluid_Dynamics}
\begin{equation}\label{eq:Pressure_Workless}
  \int_\Omega (\bm\nabla p \cdot \bm{u}) d\bm{x} =0,
\end{equation}
which indicates that pressure forces are workless through divergence-free velocity fields that are parallel to the surface; i.e., satisfying [\ref{eq:Euler_BC}]. That is, if a velocity field satisfies continuity (i.e., divergence-free) and the no-penetration boundary condition [\ref{eq:Euler_BC}], the pressure force would not contribute to the dynamics of such a field. This fact is the main reason behind vanishing the pressure force in the first step in Chorin's standard projection method for incompressible flows \cite{Chorin_Projection}; when the equation of motion is projected onto divergence-free fields, the pressure term disappears, which is based on the Helmholz-Hodge decomposition  (e.g., \cite{Geometric_Control_Fluid_Dynamics,Wu_Vortex_Dynamics_Book}): a vector $\bm{v}\in\mathbb{R}^3$ can be decomposed into a divergence-free component $\bm{u}$ that is parallel to the surface; i.e., satisfying [\ref{eq:Euler_BC}], and a curl-free component $\nabla f$ for some scalar function $f$ (i.e., $\bm{v}=\bm{u}+\bm\nabla f$). These two components are orthogonal as shown in Eq. [\ref{eq:Pressure_Workless}]. 

From the above discussion, it is clear that the pressure force is a constraint force and the dynamics of ideal fluid parcels are subject to no impressed forces\footnote{This will not be the case for a non-homogeneous, normal-flow boundary condition}. Hence, Gauss' principle of least constraint reduces to Hertz' principle of least curvature in this case. 

Considering the dynamics of an ideal fluid [\ref{eq:Euler_eq}], and labeling fluid parcels with their Lagrangian coordinates $\bm\xi$ (initial positions), we write the Appellian as 
\[ S = \int \frac{1}{2} \rho_0 \bm{a}^2 d\bm\xi, \]
where $\rho_0=\rho_0(\bm\xi)$ is the initial density. Realizing that $\rho J = \rho_0$, where $J$ is the Jacobian of the flow map \cite{Bateman_Variational_Principle}, the Appellian is rewritten in Eulerian coordinates as
\begin{equation}\label{eq:Euler_Appellian}
S = \int_\Omega \frac{1}{2} \rho \bm{a}^2 d\bm{x},
\end{equation}
which must be minimum. As such, the dynamics of an ideal fluid can be represented in the Newtonian-mechanics formulation by Eqs. [\ref{eq:Euler_eq}, \ref{eq:Euler_cont}, \ref{eq:Euler_BC}]. We present an equivalent analytical-mechanics (variational) formulation: \begin{align}
\label{eq:Euler_Hertz}
\min S = \frac{1}{2} \rho \int_\Omega  \bm{a}^2 d\bm{x}, 
\intertext{subject to continuity}
\label{eq:Euler_cont2}
\bm\nabla \cdot \bm{u} =& 0, \mathrm{ \hspace{15pt} in  \hspace{15pt} \Omega}\\
\intertext{and the no-penetration boundary condition}
\label{eq:Euler_BC2}
\bm{u}\cdot \bm{n} =& 0, \mathrm{ \hspace{15pt} on  \hspace{15pt} \delta{\Omega}},
\end{align}

\section{Dynamical Closure Condition: A Variational Theory of Lift}
Consider the standard potential flow over an airfoil (e.g., \cite{Schlichting}). It is straight forward to determine a flow field $\bm{u}_0$ that is (i) divergence-free, (ii) irrotational, and (iii) satisfies a given (possibly non-homogeneous) no-penetration boundary condition:
\begin{equation}\label{eq:Non_Homogeneous_BC}
    \bm{u}_0\cdot\bm{n}=g(\bm{x}), \;\; \mbox{on}\;\; \delta\Omega,
\end{equation}
where $g$ is a given function. This flow field is a solution of Euler's equation [\ref{eq:Euler_eq}]. Let $\bm{u}_1$ be any velocity field that is (i) divergence-free, (ii) irrotational, and (iii) satisfies a homogeneous no-penetration boundary condition [\ref{eq:Euler_BC2}]; e.g., for the cylinder, $\bm{u}_1$ can be the potential velocity field due to a unit circulation located at the center of the cylinder. Then, for any arbitrary $\Gamma$, the flow field $\bm{u}(\bm{x};\Gamma)=\bm{u}_0\bm{u}(\bm{x}) + \Gamma\bm{u}_1\bm{u}(\bm{x})$ is also a legitimate Euler's solution for the problem: it satisfies continuity, Euler's equation [\ref{eq:Euler_eq}] and the given no-penetration boundary condition [\ref{eq:Non_Homogeneous_BC}]. That is, Euler's equation does not possess a unique solution for this problem (it is a singular case where Newtonian and Lagrangian mechanics fail to determine a unique solution). In contrast, the developed variational principle is capable of providing a unique solution. In fact, the dynamics of the free-field $\Gamma\bm{u}_1$ is exactly amenable to the developed variational principle since it (i) is divergence-free, and (ii) satisfies a homogeneous no-penetration boundary condition [\ref{eq:Euler_BC2}]; that is, the pressure force is indeed orthogonal (in the sense of function spaces) to this field, as shown in Eq. [\ref{eq:Pressure_Workless}], and hence, it evolves freely (under no forces) with a minimum curvature according to Hertz' principle.


\begin{figure}
\vspace{-0.5in}
\includegraphics[width=7in]{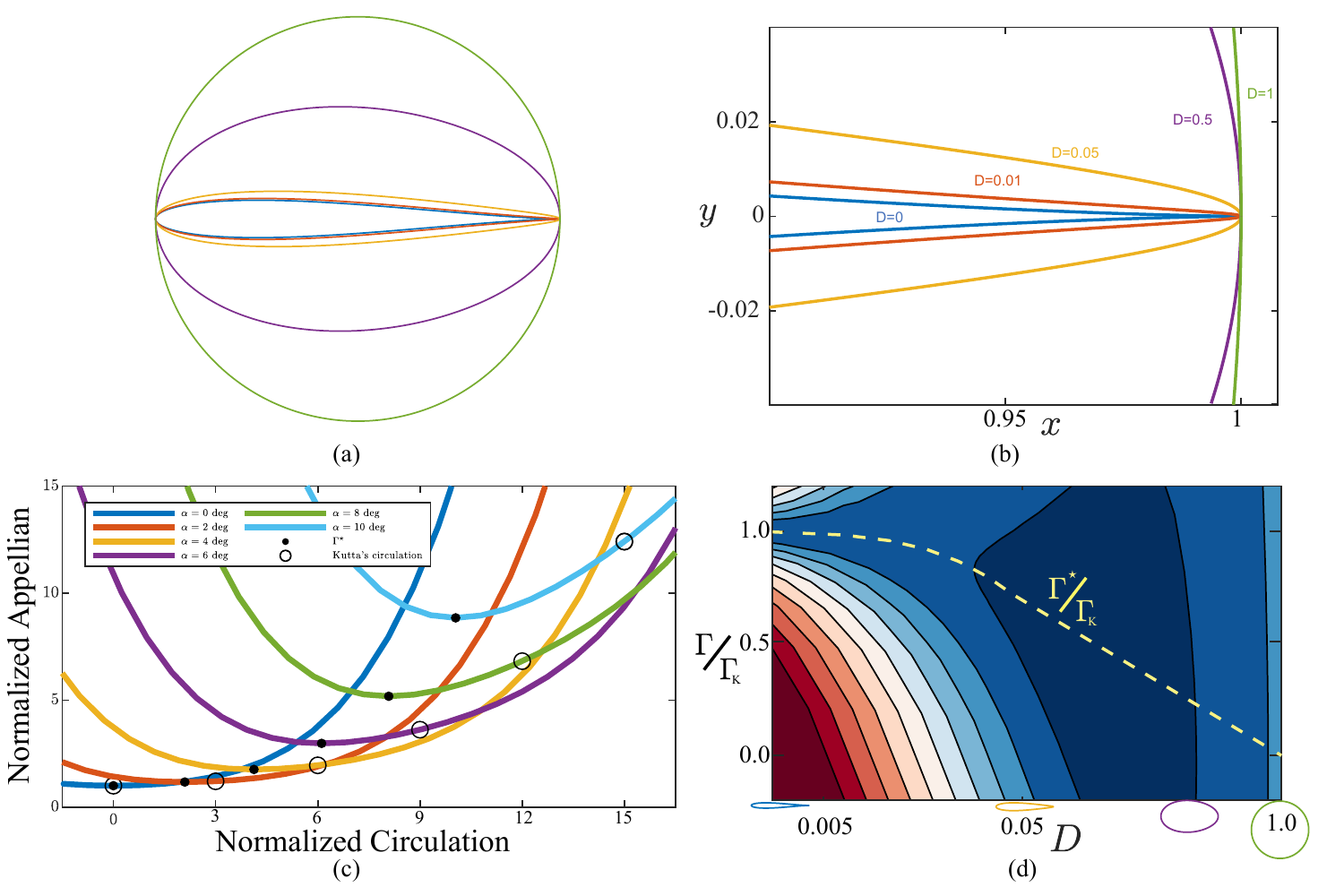}
\caption{The first row shows the effect of the shape-control parameter $D$: $D=0$ results in a a Zhukovsky airfoil with a sharp trailing edge; the larger the $D$, the smoother the trailing edge; and $D=1$ results in a smooth circular cylinder. Unlike the classical theory, the current theory is capable of predicting lift over this spectrum of shapes. The second row shows the results of the developed variational theory. Panel (c) shows the variation of the normalized Appellian $\hat{S}=\frac{S}{\rho U^4}$ with the normalized circulation $\hat\Gamma=\frac{180}{\pi}\frac{\Gamma}{4\pi Ub}$ at different angles of attack for the flow over a modified Zhukovsky airfoil with a smooth trailing edge ($D=0.05$). For each angle of attack, there is a unique value $\Gamma^*$ of circulation that minimizes the Appellian: the dynamically-correct circulation according to Hertz' principle of least curvature. Although Kutta's circulation $\Gamma_K$ is not relevant in this case of a smooth trailing edge, it is shown for comparison. Panel (d) shows contour plots of the Appellian versus the shape-control parameter $D$ and the circulation (normalized by Kutta's). Each point in the $D-\Gamma$ plane represents a flow field, not necessarily a physical flow field; Nature selects a unique flow field for each geometry (i.e., for each $D$): the one with the minimum curvature (minimum Appellian). The yellow curve presents the locus of such minimum-curvature choices. For $D=0$ (i.e., a sharp-edged airfoil, the minimizing circulation $\Gamma^*$ coincides with Kutta's $\Gamma_K$; and for $D=1$ (circular cylinder), the minimizing circulation vanishes, implying no inviscid lifting capability of this purely symmetric shape. That is, the developed theory could capture the spectrum between the two extremes: from a zero lift over a circular cylinder to the Kutta-Zhukovsky lift over a sharp-edged airfoil. And the fact that the Kutta condition is a special case of the developed inviscid theory challenges the accepted wisdom about the viscous nature of the Kutta condition. It is not a manifestation of viscous effects, but rather momentum conservation.}\label{Fig:Large_Figure}\vspace{-0.7in}
\end{figure}

Considering a steady snapshot (i.e., $\bm{a}=\bm{u}\cdot\bm\nabla \bm{u}$), we write the Appellian from [\ref{eq:Euler_Appellian}] as
\begin{equation}\label{eq:Airfoil_Appellian}
S(\Gamma) = \frac{1}{2}\rho \int_\Omega \left[\bm{u}(\bm{x};\Gamma)\cdot\bm\nabla \bm{u}(\bm{x};\Gamma)\right]^2 d\bm{x}.
\end{equation}
And the minimization principle [\ref{eq:Euler_Hertz}], derived from Hertz' principle of least curvature, yields the circulation over the airfoil as 
\begin{equation}\label{eq:Kutta_General}
\Gamma^* =\rm{argmin}\; \frac{1}{2}\rho \int_\Omega \left[\bm{u}(\bm{x};\Gamma)\cdot\bm\nabla \bm{u}(\bm{x};\Gamma)\right]^2 d\bm{x}.
\end{equation}
Equation [\ref{eq:Kutta_General}] provides a generalization of the Kutta-Zhukovsky condition that is, unlike the latter, derived from first principles: Hertz' principle of least curvature. In contrast to the classical theory, the current one is not confined to sharp-edged airfoils.


Consider a circular cylinder of radius $b$ in the $\zeta$-domain that is mapped to a modified Zhukovsky airfoil of chord length $c$ in the $z$-domain through the mapping
\[ z =  \zeta + \frac{1-D}{1+D}\frac{\delta^2}{\zeta}, \]
where $\delta$ is a constant that depends on the airfoil geometry (maximum thickness and camber) as well as the parameter $D$, which controls smoothness of the trailing edge: $D=0$ results in the classical Zhukovsky airfoil with a sharp trailing edge, and $D=1$ results in a circular cylinder, as shown in Fig. \ref{Fig:Large_Figure}(a,b). The airfoil is subject to a stream of an ideal fluid of density $\rho$ with a free stream velocity $U$ at an angle of attack $\alpha$. 

Figure \ref{Fig:Large_Figure}(c) shows the variation of the Appellian as given by Eq. [\ref{eq:Airfoil_Appellian}] and normalized by $\rho U^4$ versus the normalized circulation $\hat\Gamma=\frac{180}{\pi}\frac{\Gamma}{4\pi Ub}$ (i.e., the free parameter) at various angles of attack for the flow over a modified Zhukovsky airfoil with a trailing edge radius of 0.1\% chord length ($D=0.05$). The figure also shows Kutta's circulation $\Gamma_K=4\pi b U\sin\alpha$ (i.e., $\hat\Gamma_K\simeq\alpha^\circ$ for small angles). Note that Kutta's solution is not really applicable here. The figure shows that at a given angle of attack, the Appellian possesses a unique minimum at a specific value of the circulation---\emph{the dynamically-correct circulation according to Hertz' principle.}

Figure \ref{Fig:Large_Figure}(d) shows contours of the normalized Appellian (in logarithmic scale) in the $D$-$\Gamma$ space at $\alpha=5^\circ$ (the picture is qualitatively similar at other angles of attack). The figure also shows the locus of the minimizing circulation $\Gamma^{\star}$ (i.e., the variation of $\Gamma^{\star}$ with $D$). Interestingly, for a sharp-edged airfoil ($D=0$), the minimizing circulation coincides with Kutta's circulation (i.e., $\Gamma^{\star}=\Gamma_K$), which implies that the developed minimization principle (\ref{eq:Kutta_General}) reduces to the Kutta condition in the special case of a sharp-edged airfoil. Also, the figure shows that the smoother the trailing edge, the smaller the circulation (and lift); i.e., an airfoil with a sharp trailing edge generates larger lift than an airfoil with a smooth trailing edge at the same conditions. In fact, the figure presents the other limiting case  (a circular cylinder: $D=1$) where the classical result of the non-lifting nature of a circular cylinder in an inviscid fluid is recovered. 


\section{Discussion}
The last section presents a new theory of lift based on Hertz' variational principle of least curvature. In particular, Eq. [\ref{eq:Kutta_General}] presents the sought general fundamental principle that provides closure in potential flow based on first principles (Hertz' principle of least curvature), thereby generalizing the century-old theory of Kutta and Zhukovsky. This principle allows, for the first time, computation of lift over smooth shapes without sharp edges where the Kutta condition fails, which confirms that a sharp trailing edge is not a necessary condition for lift generation \cite{Lift_Smooth_Shapes1,Lift_Smooth_Shapes2}. That is, The new variational theory, unlike the classical one, is capable of capturing the whole spectrum between the two extremes: from zero lift over a circular cylinder to the Kutta-Zhukovsky lift over an airfoil with a sharp trailing edge, as shown in Fig. \ref{Fig:Large_Figure}(d). This behavior provides credibility to the developed theory.  

The fact that the minimization principle [\ref{eq:Kutta_General}] reduces to the Kutta condition in the special case of a sharp-edged airfoil, wedded to the fact that this principle is an inviscid principle imply that the classical Kutta-Zhukovky lift over an airfoil with a sharp edge is both computed and explained from inviscid considerations. This finding challenges the accepted wisdom about  the Kutta condition being a manifestation of viscous effects. Rather, it represents inviscid momentum conservation. That is, the circulation (and lift) is the one that satisfies momentum conservation---equivalently, it is the one that minimizes the Appellian in the language of Hertz. This result explains the several inviscid computations that converged to the Kutta-Zhukovky lift without viscosity \cite{Inviscid_Lift_CFD1,Inviscid_Lift_CFD2,New_Theory_of_Flight,Inviscid_Lift_Quantum_PRL}.


The developed theory and obtained results are intimately related to the ability of an ideal fluid (superfluid) to generate lift. So, it may be prudent to discuss the seeming contradiction between Craig and Pellam's  experimental result \cite{Inviscid_Lift_Caltech_EXperiment_PRL} and the recent computations by Musser et al. \cite{Inviscid_Lift_Quantum_PRL}, in the light of the developed theory. Craig and Pellam have experimentally studied the flow of an ideal fluid (superfluid Helium II) over a flat plate and an ellipse and found a vanishingly small lift force over these objects at non-zero angles of attack. Hence, they concluded that superfluids are non-lifting and that ``the classical viscosity boundary condition at the trailing edge (the Kutta condition) does not apply" \cite{Inviscid_Lift_Caltech_EXperiment}---the last part of this statement about the viscous nature of the Kutta condition is refuted above. 


In contrast to the above hypothesis, Musser et al. \cite{Inviscid_Lift_Quantum_PRL} recently performed quantum simulations of the Gross-Pitaevskii equation governing the flow dynamics over an airfoil in a superfluid. Their simulations show a quantized version of the Kutta-Zhukovsky lift despite the lack of viscosity in their simulations. However, the continuum hypothesis may not be applicable in their ultra-small-scale simulations. 


From the above discussion, there seems to be contradiction between Craig and Pellam's experiment \cite{Inviscid_Lift_Caltech_EXperiment_PRL} and the computations of Musser et al. \cite{Inviscid_Lift_Quantum_PRL}: the former deceitfully implies that superfluids are non-lifting and the Kutta condition is a viscous condition (i.e., the physics is discontinuous in the limit), while the latter shows the ability of a superfluid to generate the Kutta-Zhukovsky lift similar to a viscous flow of a vanishingly small viscosity (i.e., the physics is continuous in the limit). In the light of the developed theory, we find no contradiction between the two results. In fact, Craig and Pellam conducted their experiment on symmetric shapes (a flat plate and an ellipse) for which the current theory predicts no lift, which confirms their experimental findings (but not their conclusions). Also, the current theory predicts Kutta-Zhukovsky lift over an airfoil with a sharp trailing edge; and a smaller but non-zero lift over a smooth airfoil (or asymmetric shape), which confirms Musser et al. computational simulations \cite{Inviscid_Lift_Quantum_PRL}. Therefore, it is concluded that a purely symmetric shape in a superfluid is non-lifting; however, any asymmetry would grant some lifting capability. The current theory, as well as Musser et al. quantum simulations \cite{Inviscid_Lift_Quantum_PRL}, invoke an experimental study of the flow of a superfluid  (Helium II), similar to Craig and Pellam's \cite{Inviscid_Lift_Caltech_EXperiment_PRL}, but over a traditional airfoil shape (i.e., asymmetric shape).

\begin{wrapfigure}{l}{0.50\textwidth}
 \begin{center}
 \includegraphics[width=8cm]{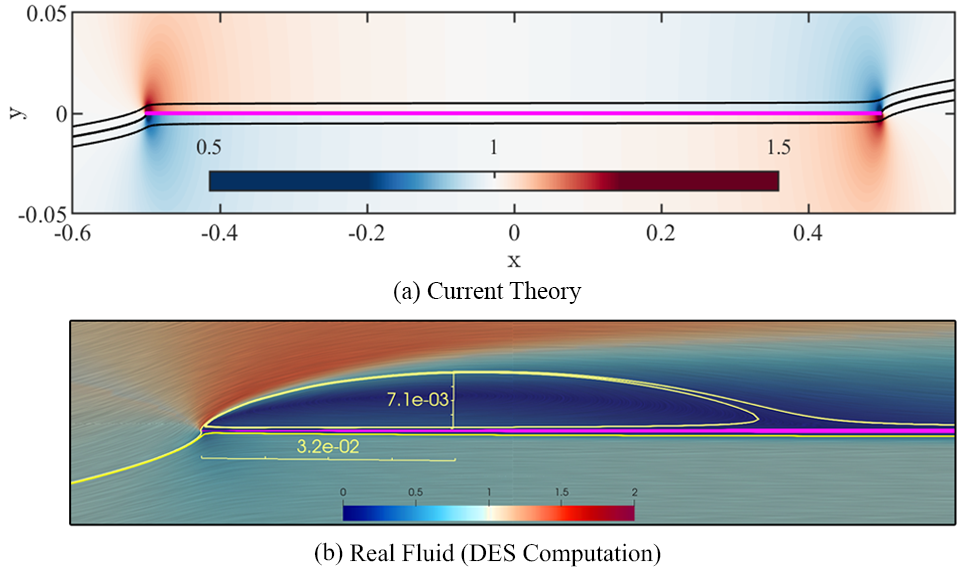}
 \caption{Flow velocity, $u/U$, of a superfluid (using the current theory) and a real fluid (using Detached Eddy Simulations) over a flat plate at $\alpha=2^\circ$.}
\label{Fig:Flat_Plate_Ideal_Flow}\vspace{-0.2in}
 \end{center}
\end{wrapfigure}

The fact that purely symmetric shapes are non-lifting in an ideal (non-dissipative) fluid is physically plausible, as shown in Fig. \ref{Fig:Flat_Plate_Ideal_Flow}(a) presenting the flow of an ideal fluid over a flat plate as predicted by the current theory. These flows must be  reversible to conserve entropy in the dissipation-free environment. Therefore, for reversibility and because of symmetry, one should not be able to tell whether the flow field, shown in Fig. \ref{Fig:Flat_Plate_Ideal_Flow}(a), is for a free stream coming from the left at a positive $\alpha$ or from the right at a negative $\alpha$. Clearly, this symmetric solution is non-lifting.

For these symmetric shapes, viscosity is important to enable the \textit{weak} lift over these bodies: the slight change of the effective body shape due to boundary layer destroys symmetry; the outer inviscid flow over the modified asymmetric body is now lifting. The flat plate represents an extreme case in this regard: the effect of viscosity is significant due to the singular nature of the corresponding ideal flow field. Viscosity leads to separation at the leading edge even at small angles of attack, as shown in Fig. \ref{Fig:Flat_Plate_Ideal_Flow}(b), which presents our Detached Eddy Simulation (DES) of the flow over a flat plate at a Reynolds number of 500,000 and $\alpha=2^\circ$, also see Van Dyke's Album of Fluid Motion \cite{VanDykeAlbum}. A leading-edge separation bubble is clearly seen on the top of the flat plate in the immediate vicinity of the leading edge. The outer inviscid flow field outside the separation bubble (i.e., over the modified flat plate: a flat plate with a \textit{naturally} rounded nose) is lifting; the minimization principle [\ref{eq:Kutta_General}], which is equivalent to the Kutta condition in this case of a sharp-trailing edge, results in a circulation $\Gamma^{\star}$ that is close to Kutta's.

\noindent The developed theory is expected to deepen our understanding of one of the most fundamental concepts in aerodynamics: lift generation over an airfoil. 



Finally, it may be prudent to provide a simple explanation for lift generation in the light of the obtained results and presented discussion---an explanation that does not invoke complex unsteady phenomena to illustrate how/why a particular steady lift value is attained. It is found that lift is due to the triplet [\ref{eq:Euler_eq},\ref{eq:Euler_cont},\ref{eq:Euler_BC}], equivalently [\ref{eq:Euler_Hertz},\ref{eq:Euler_cont2},\ref{eq:Euler_BC2}]. In words, lift is due to (i) momentum conservation, (ii) continuity (the fluid has to fill the given space), and (iii) the body's hard constraint (i.e., presence of the body inside the fluid), which forces the flow to move around the body without creating void to maintain (ii). That is, lift is due to conservation of momentum between the two equilibrium states of the freestream and the curved flow around the body. In layman terms, the presence of the body inside the fluid forces fluid particles to go along a curved path; i.e., it provides the necessary centripetal force for the  curved path. And the fluid responds back by an equal amount of force in the opposite direction according to Newton's third law---this reaction is the lift force. This explanation is similar to Hoffren's \cite{Lift_Quest} and to some extent to Babinsky's \cite{Babinsky_Lift_Generation}; the current paper provides the sought mathematical proof. 
Inventors and analysts alike may find it intriguing that a ``Dry Water'' model \cite{FeynmanVol2} provides the main physical mechanism behind one of most fundamental problems in fluid mechanics.

\bibliography{Fluid_Dynamics_References,Aeronautical_Engineering_References,Dynamics_Control_References,Geometric_Control_References,Viscous_Corrections_Ref}

\end{document}